\documentclass[sn-mathphys-num]{sn-jnl}
\usepackage{graphicx}%
\usepackage{multirow}%
\usepackage{amsmath,amssymb,amsfonts}%
\usepackage{amsthm}%
\usepackage{mathrsfs}%
\usepackage[title]{appendix}%
\usepackage{xcolor}%
\usepackage{textcomp}%
\usepackage{manyfoot}%
\usepackage{booktabs}%
\usepackage{algorithm}%
\usepackage{algorithmicx}%
\usepackage{algpseudocode}%
\usepackage{listings}
\usepackage{bm}

\makeatletter

\theoremstyle{thmstyleone}%
\theoremstyle{thmstyletwo}%

\theoremstyle{thmstylethree}%

\makeatother

\usepackage{babel}
\begin{document}
\title{Fast-moving electrostatic solitons in a plasma with turbulence heating}

\author[1]{\fnm{Mridusmita} \sur{Das}}

\author[2]{\fnm{Murchana} \sur{Khusroo}}

\author*[1]{\fnm{Madhurjya P.} \sur{Bora}}\email{mpbora@gauhati.ac.in}

\affil*[1]{\orgdiv{Department of Physics}, \orgname{Gauhati University}, \city{Guwahati},
\postcode{781014}, \country{India}}

\affil[2]{\orgdiv{Department of Physics}, \orgname{University of Science \& Technology Meghalaya},
\city{Ri-Bhoi}, \postcode{793101}, \country{India}}
\abstract{{In this work, it is shown} that electrostatic solitons in a plasma with turbulent
heating of the electrons through an accelerating electric field can
form with very high velocities, reaching up to several order of magnitudes
larger than the equilibrium ion-sound speed. The possible parameter
regime, where this work may be relevant, can be found in the so-called
``dead zones'' of a protoplanetary disk. {Though these zones are stable
to magnetorotational instability, the resultant turbulence can in fact heat the electrons making
them follow a highly non-Maxwellian velocity distribution.} We show
that these fast-moving solitons can reach very high velocities. With
electron velocity distribution described by the Davydov distribution
function, we argue that
these solitons can be an effective mechanism for energy equilibration
in such a situation through soliton decay and radiation.}
\keywords{soliton, turbulence}
\maketitle

\section{Introduction}

In this work, we consider the plasma environment of a protoplanetary
disk and the formation of electrostatic solitons in such a plasma.
In these plasmas, there can be turbulence-driven electric field causing
random heating of the electrons pushing the electrons to have a highly
non-Maxwellian distribution. Protoplanetary disks (PPD) are circumstellar
disks around young stars, which usually precede the formation of planets.
These disks are primarily made up of gas, dust, and debris, which
eventually break up to give rise to planets and similar smaller orbiting
objects around the parent star \citep{pringle,eric,russel,armitage}.
They are formed around the equatorial plane of the star with radii
ranging to several AUs. They are also quite cold objects with temperatures
gradually falling from about $1000\,\textrm{K}$ to about $100\,\textrm{K}$
as one goes out from the inner to the outer regions of the disk \citep{lesur,taka}.
Apart from being weakly ionized plasmas, the environment of a PPD
also contains a considerable amount of dust grains. The electrostatic
interaction between the dust grains and the plasma can give rise to
various plasma processes which can considerably affect the resulting
nonlinear interactions \citep{vitaly}. In recent years, we are observing
a renewed interest in studying PPDs, primarily due to their important
roles in the formation of planets and recent advances in space technology
which have led to the discovery of several exoplanets. However, the
exact mechanisms which eventually break down a PPD leading to planet
formation, can still be debated about, as there are several competing
processes which can simultaneously develop to form a quite complex
scenario. Although the magnetorotational instability (MRI) \citep{Balbus-1991,Hawley-1995}
is believed to be a major candidate for the reorganization of disk
material, the presence of dust grains creates the so-called ``dead
zones'' where MRI can no longer operate. In contrast to the dead-zones,
there is an ``active zone'', which is a turbulent envelope of plasma
that surrounds the dead zone \citep{gammie}. We should also note
that the ambient large-scale magnetic field in a PPD is actually very
small, about a Gauss at $\sim1\textrm{AU}$ to a few milli-Gauss,
as one moves out to the extreme outer region \citep{wardle}. In the
outer region, the magnetic field is so weak that no observational
evidence of the magnetic field could be detected through related phenomena
such as the Zeeman effect \citep{lesur}.

The coupling between a moving plasma and magnetic field through MRI-driven
turbulence can produce a finite electric field, even in a dead zone,
which can be quite strong. In a weakly ionized plasma such as in the
outer region of a PPD, such an electric field, if sufficiently strong,
can produce random motion and cause heating \citep{Inutsuka-2005}.
The acceleration of the electrons by this electric field is balanced
due to the collisional loss of energy and momentum with the neutrals.
Though these collisions can both be elastic and inelastic, the loss
of energy due to inelastic collisions can be largely ignored for energies
$\lesssim1\,\textrm{eV}$ \citep{Okuzumi-2015}. This approximation
allows the electron velocity distribution to be written in a form
which is known as \emph{Davydov} distribution \citep{golant,Lifshitz-1981,Davydov-1935,Okuzumi-2015}.
On the other hand, the ion distribution function still remains largely
Maxwellian \citep{Okuzumi-2015}. If we can ensure electrical quasi-neutrality
(see the discussion on the validity of quasi-neutrality in {Section
3.1)} in such a plasma with the relevant timescale (ion-acoustic),
we can describe it with the ion fluid equations, stationary dust grains,
and electrons with Davydov distribution function. The charged dust
grains affect the plasma dynamics by depleting the electrons and by
maintaining a static electrical background.

Coming to the subject of nonlinear electrostatic structures like solitons,
envelope solitons, and shock waves, we note that they are of very
special interest to plasmas. Usually, in both space and laboratory
plasmas, the turbulence becomes high enough such that nonlinear and
dispersion effects become comparable and localized electrostatic structures
like solitons and envelope solitons start to form \citep{galeev}.
However, it is now well known that plasma solitons are almost always
unstable \citep{stability} and can be an effective mechanism for
energy equilibration through processes like soliton radiation, where
a soliton effectively \emph{radiates} ion-acoustic waves, which in
turn causes the soliton amplitude and velocity to decrease \citep{schamel,rad}.
In this work, we show that in a situation, where turbulence-driven
heating can change the effective electron distribution function, large-amplitude
electrostatic solitons can form and can reach very high velocities
up to several orders of magnitude higher than the equilibrium ion-sound
velocity. These large, fast-moving solitons can then be very effective
pathways for energy transfer in such a plasma. As discussed above,
these solitons can form in the ``dead zones'' and transfer energy
outward. In Section {2}, we briefly discuss the theory of MRI turbulence
and plasma heating \citep{Okuzumi-2015}. We also discuss the relevant
parameter regime where our analysis can be useful, toward the end
of this section. In Section {3} and the subsequent subsection, we
formulate the electron velocity distribution and the electron density.
We discuss the nonlinear electrostatic waves in Section {4} and also
discuss the possible effects of dust-charging. In Section {5}, we briefly
discuss the {behaviour} of nonlinear electrostatic waves when no dust
is present. In Section {6}, we discuss the full implications of dust
effects and discuss the most important findings of this analysis.
In this section, we also outline a mathematical procedure, which can
be used to make the analytical interpretation of these waves easier.
In Section {7}, we conclude.

\section{Debye-scale structures in a turbulent plasma}

\textcolor{black}{Turbulence is one of the most common and universal
nonlinear phenomena observed in naturally occurring fluids. In electrically
active fluid like plasma, electrostatic and electromagnetic turbulence
are routinely observed in space plasmas such as solar wind, plasmas
in the magnetosphere, interplanetary plasmas, as well as in laboratory
plasmas \citep{turb}. While different manifestations of nonlinear phenomena
like shock waves (both dissipation and dissipation-less), solitons,
and double layers can occur independent of each other, a turbulent
regime can give rise to all these under suitable environment. In general,
a turbulent flow is characterised by the so-called inertial subrange
where large-scale eddies feed energy into successively smaller eddies,
before the energy is dissipated away through viscous drag. Naturally,
viscosity does not play any significant role during the inertial subrange.
This inertial subrange can be compared to what is known as local thermodynamic
equilibrium (LTE) in the context of energy flow in stellar interiors
where an `equilibrium' is achieved despite continuous energy flow
from the centre of a star to its outer areas maintained at a constant
gradient of temperature \citep{rubinstein,bohm}.}

\textcolor{black}{In fact, appearances of coherent structures in turbulent
flow have been intriguing scientists for a long time. A very detailed
overview of solitons in a turbulent flow can be found in the review
article by Levich \citep{levich2,levich1}. It is now believed that
accumulation of energy at large scales is intuitively favourable to
the birth of coherent structures in a turbulent environment. There
are also plenty of observational evidences where soliton-like coherent
structures (SCS) are observed in low-speed turbulent boundary layers
\citep{lee1,lee2,lee3,jiang}, which are believed to be results of
streak instability and hairpin vortex. Other examples of coherent
structure such as solitons in a turbulent flow include vortex solitary
waves in a rotating turbulent flow \citep{hopfinger}. As far as space
plasmas are concerned, SCS are routinely observed in the form of ion
and electron holes. A very recent observation of such example through
Parker Solar Probe (PSP) is reported by Mozer et al. \citep{mozer}
in near-sun solar wind plasmas, though such observations in ion-acoustic
regime have also been reported earlier \citep{temerin}. Coherent
structures are also reported to be seen in fusion plasmas \citep{zweben1,zweben,surko,meiss-book},
where imaging of the edge region of tokamaks shows localized filaments
\citep{zweben1,zweben}. In a charged fluid like plasma, bipolar electric
fields are routinely observed within a turbulent regime in magnetospheric
and solar wind plasmas indicating development of Debye-scale potential
structures. Recent observations of Magnetospheric Multiscale (MMS)
spacecraft of bipolar electrostatic structures indicates that these
structures are formed through turbulence, driven by Buneman instability
which provides energy to the accelerating ions in Earth's bow shock
regions \citep{debye-scale,mms}. These bipolar electric fields are
results of an electrostatic potential hump or electrostatic solitons
arising out of this turbulent equilibrium. Very recent numerical simulation
also indicates that external charged debris can cause localised solitons
known as pinned solitons formed out of ion-ion counter-streaming turbulence
\citep{mrid}. In any case, we can very well consider the formation
of electrostatic potential hump in an otherwise turbulent flow. Mathematically,
this is equivalent to considering an electrostatic perturbation in
an equilibrium background. We argue that this must happen within the
energy spectrum of the inertial subrange of the turbulence as viscosity
does not seem to play a significant role in formation of these structures.}

\textcolor{black}{We now recall that an electrostatic potential hump
or electrostatic soliton can be mathematically described by a pseudo-potential
$V(\phi)$ through the well-known Sagdeev potential equation
\begin{equation}
\frac{d^{2}\phi}{dx^{2}}=f(\phi)=-\frac{dV}{d\phi},
\end{equation}
which is basically a re-framed version of Poisson's equation for electrostatic
potential $\phi$. In the weak nonlinearity limit, this equation can
be reduced to a Korteweg-de Vries (KdV) equation \citep{kdv} which
admits the well-known `${\rm sech}$' solution depicting a soliton.
We should note here that while the KdV equation is fully integrable,
another widely used, non-integrable equation which also admit a `${\rm sech}$'
solitary wave solution is the regularized-long-wave (RLW) equation,
first proposed by Peregrine in 1966 \citep{perigrene}. In fact, in
the small wave number limit the RLW equation reduces to a KdV equation.
We, in this work, however describe our solitary structure through
the pseudo-potential equation.}

\section{MRI turbulence and plasma heating}

In this section, we briefly review the physics of local heating of
the electrons in a weakly ionized plasma due to the strong electric
field generated through MRI turbulence in an environment similar to
a PPD, as discussed in {details} by Okuzumi and Inutsuka \citep{Okuzumi-2015}.

We consider an $e$-$i$ plasma with a considerable presence of neutrals.
As with any plasma, we expect the plasma particles to thermalize with
the neutrals when there is no external force field present in the
plasma. The situation becomes vastly different in the presence of
an external electric field. Though both electrons and ions get accelerated
in the opposite directions by the field, {effective} electrons are
heavily accelerated due to their high mobility. Besides the energy
transfer during binary collisions between the electrons and neutrals
is quite low \citep{Okuzumi-2015}. It can be proved that in this
scenario \citep{Druyvesteyn-1940,Lifshitz-1981}, one can attain an
equilibrium if the magnitude of the electric field $E$ exceeds a
certain critical value
\begin{equation}
E_{{\rm crit}}=\left(\frac{T_{n}}{el_{e}}\right)\sqrt{\frac{6m_{e}}{m_{n}}},
\end{equation}
where $l_{e}=(n_{n}\sigma_{en})^{-1}$ is the electron mean free path
and $\sigma_{en}$ is the electron-neutral momentum-transfer cross
section \citep{murchana} and the `$n$' subscript denotes the neutrals.
In this case the random thermal energy of the electrons heavily overwhelms
that of the neutrals $v_{{\rm th}e}\gg\sqrt{T_{n}/m_{n}}$. For a
hydrogen-rich environment with $\sigma_{en}\sim10^{-15}\,{\rm cm}^{2}$
at electron energy $<10\,{\rm eV}$, we have \citep{frost,yoon}
\begin{equation}
E_{{\rm crit}}\sim10^{-9}T_{100}n_{12}\,{\rm esu\,cm^{-2}}.
\end{equation}
In the above relation, $T_{100}$ is the temperature measured in terms
of $100\,{\rm K}$ and $n_{12}$ is the neutral number density measured
in terms of $10^{12}\,{\rm cm^{-3}}$. It can be shown that \citep{Okuzumi-2015}
\begin{equation}
\frac{E_{{\rm MRI}}}{E_{{\rm crit}}}\approx\frac{200}{\Lambda}\left(\frac{100}{\beta_{z}}\right)n_{12}^{-1/2},
\end{equation}
where $E_{\textrm{MRI}}$ is the strength of the electric field, generated
due to MRI turbulence in a situation like that of a PPD. It should
be noted that the above expression is independent of the gas temperature
$T$, where
\begin{equation}
\Lambda=\frac{v_{Az}^{2}}{\eta\Omega}
\end{equation}
is the Elsasser number \citep{elsasser} -- the ratio of the kinetic
energy to the Coriolis energy, $\beta_{z}$ is the plasma $\beta$
of the vertical magnetic field, and $v_{Az}$ is the corresponding
Alfv\'en velocity. An upper limit for $\Lambda$ can be established
by noting the fact that for electron heating, one must have $E_{{\rm MRI}}>E_{{\rm crit}}$.
However, for sustaining the MRI-driven turbulence, $E_{{\rm MRI}}$
has to be sufficiently strong, from which we have the condition that
$\Lambda>\Lambda_{{\rm crit}}$. Naturally, for a sustained MRI-driven
turbulence heating of the electrons, we must have \citep{Okuzumi-2015},
\begin{equation}
\Lambda_{{\rm crit}}\lesssim\Lambda\lesssim200\left(\frac{100}{\beta_{z}}\right)n_{12}^{-1/2}.
\end{equation}
We note in a typical situation of a PPD, $\Lambda\sim0.1-1$ and $n_{12}\sim10^{2}-10^{6}$
with $\beta_{z}\sim100-1000$, and the above condition can be readily
satisfied.

\subsection{Plasma parameters and validity}

It is necessary to validate the parameter space and validate the assumptions
that we make in subsequent sections. PPDs are weakly ionized structures
with considerable presence of neutrals and dust grains. There is also
a considerable amount of ion-neutral and electron-neutral collisions.
Considering the characteristic scale length and timescale, we can
now estimate the basic plasma parameters for this plasma, which are
Debye length $(\lambda_{D})$, plasma frequencies $(\omega_{p,j})$
and mean collision times $(\tau_{j})$ for different species, $j=i,e,d$
for ions, electrons, and dust particles. The Debye length can be parameterized
as \citep{lesur}
\begin{equation}
\lambda_{D}=\frac{3}{10}\xi_{-13}^{-1/2}R_{\textrm{AU}}^{7/8}(1+Z)^{-1/2}\,\textrm{m},
\end{equation}
where $\xi_{-13}$ is the ionization fraction $n_{-}/n_{n}$, the
ratio of free negative charge carriers $n_{-}$ and neutrals $n_{n}$,
measured in the units of $10^{-13}$, $R_{\textrm{AU }}$ is the radial
distance from the center of the PPD in the units of AU, and $Z$ is
the ion charge number. The quantity $\xi_{-13}$ is a highly uncertain
quantity, which can be estimated to $\sim1-10^{-3}$. With $Z=1$
and $R_{\textrm{AU}}\sim1$, we have $\lambda_{D}\sim1-30\,{\rm m}$
\citep{sano,lesur}, which is clearly {quite smaller} than the relevant
scale lengths in question. Apparently, the number of particles in
the Debye sphere $\sim10^{6}\gg1$. So, these objects can still be
considered very much in the plasma regime even though they have a
very low ionization and temperature \citep{lesur}. This approximation
can of course {break down} at higher charge number $Z\gtrsim10^{3}$,
which is however prohibitively large \citep{wardle}.

Coming to the question of quasi-neutrality, which is of course assumed
throughout this work, it is important to justify its validity region.
We consider the electron and ion plasma frequencies, which can be
parameterized as \citep{lesur}
\begin{eqnarray}
\omega_{p,e} & = & 2.2\times10^{5}\xi_{-13}^{1/2}R_{\textrm{AU}}^{-9/8}\,\textrm{s}^{-1},\\
\omega_{p,i} & = & 9.3\times10^{2}\xi_{-13}^{1/2}R_{\textrm{AU}}^{-9/8}\,\textrm{s}^{-1},
\end{eqnarray}
and the corresponding average mean collision times
\begin{eqnarray}
\tau_{s,e} & = & 6.7\times10^{-7}R_{\textrm{AU}}^{9/4}\,\textrm{s},\\
\tau_{s,i} & = & 4.9\times10^{-5}R_{\textrm{AU}}^{9/4}\,\textrm{s}.
\end{eqnarray}
With the relevant parameters for PPD, one can show that $10^{-2}<\omega_{p}\tau_{s}<1$,
and it can be concluded that electrical quasi-neutrality is recovered
on timescales shorter than a second \citep{lesur}. So, we can safely
conclude that quasi-neutrality is maintained by electrons and neutrals,
at least, in the ion-acoustic timescale.

The probable region of the PPD, where this work might be of relevance
is toward the outer edge of the disk where the plasma is only weakly
ionized and there exists a so-called MRI-stable ``dead zone'' \citep{lesur,Okuzumi-2015}.
However, it has been shown that the MRI-driven turbulence can produce
an electric field in the neutral co-moving frame when ionization is
low. In these regions the large-scale magnetic field is quite weak
$\sim10\,{\rm mG}$ \citep{lesur}
\begin{equation}
B\sim12R_{\textrm{AU}}^{-11/8}\beta^{-1/2}\,\textrm{G},
\end{equation}
so that the thermal pressure dominates over magnetic pressure, resulting
a very high plasma $\beta\gg1$, which the ratio of the thermal pressure
to that of the magnetic field. From which, we see that the electron
and ion gyro-radii $\gg\lambda_{d}$ and a localized electrostatic
structure can form. 

\section{Electron distribution and heating}

\textcolor{black}{Let us now consider the Davydov velocity distribution
function for the electrons. This distribution function can be considered
{as} an equilibrium distribution function in a turbulent environment with
the assumption that fluctuations within the distribution function
itself can be considered to be a part of the equilibrium distribution
function provided the relaxation timescale for these fluctuations
is considerably less that kinetic timescale on which the distribution
function itself may evolve \citep{Davydov-1935,Lifshitz-1981,Okuzumi-2015,murchana}.}

\textcolor{black}{This distribution function can be written as \citep{Davydov-1935}
\begin{equation}
f_{e}\left(\boldsymbol{E},\boldsymbol{v}\right)=\left(1-\frac{eEl}{T}\frac{\epsilon\hat{\bm{E}}.\hat{\bm{v}}}{\epsilon+\chi T}\right)f_{e0}(E,v),\label{eq:davydov-1}
\end{equation}
where $\epsilon=1/2({m_e}v^{2})$ is the kinetic energy of the electrons
having velocity $\bm{v}$ and $\chi$ is a dimensionless quantity
defined as the square-root of the ratio of the electric field $E$
to its critical value
\begin{equation}
\chi=\left(\frac{E}{E_{{\rm crit}}}\right)^{2}.\label{eq:chi-1}
\end{equation}
In Eq.(\ref{eq:davydov-1}), $\hat{\bm{E}}=\bm{E}/E$, $\hat{\bm{v}}=\bm{v}/v$,
and
\begin{equation}
f_{e0}=\left(\frac{m_e}{2\pi T}\right)^{3/2}\frac{(\epsilon/T+\chi)^{\chi}}{{\cal W}(\chi)}e^{-\epsilon/T},\label{eq:maxwellchi-1}
\end{equation}
where
\begin{equation}
{\cal W}\left(\chi\right)=\chi^{3/2+\chi}U\left(\frac{3}{2},\frac{5}{2}+\chi,\chi\right).
\end{equation}
The functions $U(x,y,z)$ is the confluent hypergeometric function
of the second kind \citep{Okuzumi-2015}. As we note, $f_{e0}(E,v)$
is the symmetric part of $f_{e}\left(\boldsymbol{E},\boldsymbol{v}\right)$
that depends on the magnitudes of $\boldsymbol{E}$ and $\boldsymbol{v}$
but not on the angle between them $\cos^{-1}(\hat{\bm{E}}\cdot\hat{\bm{v}})$.
In the limit of weak electric field $(E\ll E_{{\rm crit}})$, $f_{e0}$
reduces to the usual Maxwellian distribution,
\begin{equation}
f_{e0}^{M}=\left(\frac{m_e}{2\pi T}\right)^{3/2}e^{-\epsilon/T}.\label{eq:maxwell}
\end{equation}
In the limit of a strong electric field $(E\gg E_{{\rm crit}})$,
$f_{e0}$ reduces to the Druyvesteyn distribution function \citep{Druyvesteyn-1940},
\begin{equation}
f_{e0}^{D}=\frac{1}{\pi\Gamma(3/4)}\left[\frac{3m_{e}^{3}}{4m_{n}(eEl_{e})^{2}}\right]^{3/4}\exp\left[-\frac{3m_{e}\epsilon^{2}}{m_{n}(eEl)^{2}}\right].
\end{equation}
}

\textcolor{black}{We now introduce a turbulence-generated electrostatic
potential $\phi$, expressed in terms of the equivalent electrostatic
energy $\epsilon_{p}=e\phi$. The distribution function in Eq.(\ref{eq:davydov-1})
becomes,}
\begin{equation}
f_{e}\left(\boldsymbol{E},\boldsymbol{v}\right)=\left[1-\frac{eEl_e}{T}\frac{(\epsilon-\epsilon_{p})\hat{\bm{E}}.\hat{\bm{v}}}{\epsilon-\epsilon_{p}+\chi T}\right]f_{e0},\label{eq:perturbeddavydov}
\end{equation}
{where, the} symmetric and the asymmetric parts of the distribution
function can be written as \citep{murchana},
\begin{eqnarray}
f_{e0} & = & \left(\frac{m_e}{2\pi T}\right)^{3/2}\frac{\left[(\epsilon-\epsilon_{p})/T+\chi\right]^{\chi}}{{\cal W}\left(\chi\right)}\,e^{-(\epsilon-\epsilon_{p})/T},\label{eq:perturbedsymmetric}\\
f_{\textrm{asym}}\left(\boldsymbol{E},\boldsymbol{v}\right) & = & -\frac{eEl_e}{T}\frac{(\epsilon-\epsilon_{p})\hat{\bm{E}}.\hat{\bm{v}}}{\epsilon-\epsilon_{p}+\chi T}f_{e0}
\end{eqnarray}
\textcolor{black}{The corresponding electron density can be calculated
by taking the zeroth order moment of the distribution function,}
\begin{equation}
n_{e}=\intop_{-\infty}^{+\infty}f_{e}(\boldsymbol{E},\boldsymbol{v})\,d^{3}v\label{eq:density}
\end{equation}
The above integral {in  spherical} coordinate can be written as \citep{chen-1974},
\begin{equation}
n_{e}=\int_{0}^{2\pi}\int_{0}^{\pi}\int_{0}^{\infty}v^{2}\sin\theta f_{e}\left(\boldsymbol{E},v\right)\,d\varphi d\theta dv\label{eq:densityspherical}
\end{equation}
where, the volume element of each spherical shell is $4\pi\bm{v}^{2}d\bm{v}$.
However, noting that $\hat{\bm{E}}.\hat{\bm{v}}=\cos\theta$, we can
see that the integration of the asymmetric part of the distribution
function evaluates to zero, so that the electron density is entirely
due to the symmetric part of the distribution function
\begin{eqnarray}
n_{e} & = & \intop_{0}^{\infty}4\pi v^{2}f_{e0}\,dv,\nonumber \\
 & = & e^{\epsilon_{p}}\frac{U\left(-\chi,-{\displaystyle \frac{1}{2}}-\chi,-\epsilon_{p}+\chi\right)}{U\left(-\chi,-{\displaystyle \frac{1}{2}}-\chi,\chi\right)},\quad\chi>\epsilon_{p}\label{eq:totaldensity}
\end{eqnarray}
where we have re-scaled the energies $(\epsilon,\epsilon_{p})\to(\epsilon,\epsilon_{p})/T$.
The effect of plasma heating remains in the form of change of magnitude
of electron momenta. Also note that the electron density is normalized
by its equilibrium value, so that $n_{e}\to1$ when $\epsilon_{p}\equiv\phi\to0$.

\subsection{\textcolor{black}{Linear dispersion}}

\begin{figure}
\begin{centering}
\textcolor{black}{\includegraphics[clip,width=0.5\textwidth]{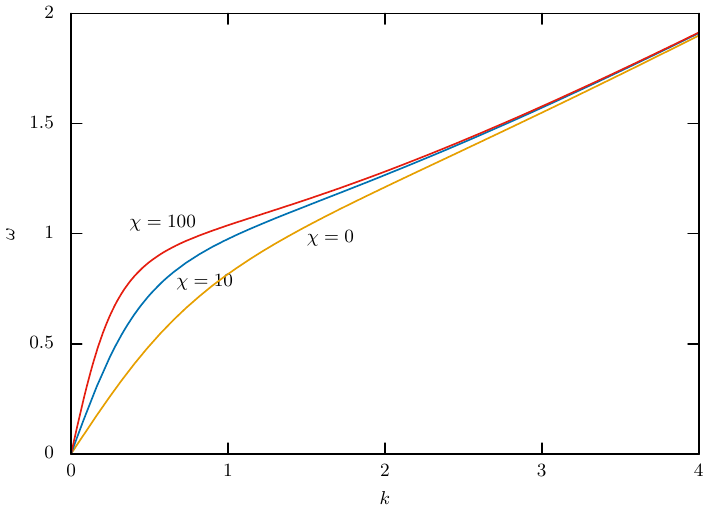}}
\par\end{centering}
\textcolor{black}{\caption{\protect\label{fig:The-linear-ion-acoustic}The linear ion-acoustic
dispersion curves for Davydov electrons. All the curves are with $\gamma=5/3$
and $\sigma=0.1$.}
}
\end{figure}

\textcolor{black}{In this subsection, we very briefly review the effect
of Davydov electrons \citep{golant,Lifshitz-1981,Davydov-1935} on the linear ion-acoustic dispersion relation. The relevant
equations in this case are ion continuity and momentum equations and
Poisson equation \citep{chen-1974}, which can be written in the dimensionless form as
\begin{eqnarray}
\frac{\partial n_{i}}{\partial t}+\nabla\cdot(n_{i}\bm{u}_{i}) & = & 0,\\
n_{i}\frac{d\bm{u}_{i}}{dt} & = & -\nabla\phi-\nabla p_{i},\\
\nabla^{2}\phi & = & n_{i}-n_{e},
\end{eqnarray}
where $p_{i}\propto n_{i}^{\gamma}$ is the ion equation of state
with $\gamma$ as the polytropic index and $\phi$ is the first order
turbulence-generated electrostatic potential. The electrons can be
considered to be inertial-less and is expressed through Eq.(\ref{eq:totaldensity}).
Without loss of any generality, we can use a frame of reference where
the equilibrium ion velocity is zero. Expressing a linear perturbation
of the form $f\to f_{0}+f_{1}$ where $f_{0,1}$ are the equilibrium
and first-order perturbed quantities with $f_{1}\sim e^{-i\omega t+i\bm{k}\cdot\bm{r}}$.
The first order perturbed quantities are represented by $f_{1}=(n_{i1},n_{e1},u_{i1},p_{i1},\phi\equiv\phi_{1})$.
Following the standard procedure, the linear dispersion relation can
be expressed
\begin{equation}
\omega=k\left(\sigma\gamma+\frac{1}{{\cal A}+k^{2}}\right)^{1/2},
\end{equation}
where Eq.(\ref{eq:totaldensity}) {is  linearised} as
\begin{equation}
n_{e1}={\cal A}\phi_{1},
\end{equation}
with
\begin{equation}
{\cal A}=1-\frac{U\left({\displaystyle \frac{3}{2}},{\displaystyle \frac{3}{2}}+\chi,\chi\right)}{U\left({\displaystyle \frac{3}{2}},{\displaystyle \frac{5}{2}}+\chi,\chi\right)}.
\end{equation}
In the above equations, we have normalised the densities by their
respective equilibrium values, $\phi_{1}$ by $T_{e}/e$, $\omega$
by $c_{s}/\lambda_{D}$, and $k$ by $\lambda_{D}^{-1}$. $\lambda_{D}$
is the electron Debye length and $\sigma=T_{i}/T_{e}$. We can easily
see that in the limit $\chi\to0$, ${\cal A}\to1$, the electrons
become Boltzmannian and we recover the usual ion-acoustic dispersion
relation.}

\textcolor{black}{In Fig.\ref{fig:The-linear-ion-acoustic}, we have
shown the linear dispersion curves for Davydov electrons. As can be
seen, a higher $\chi$ makes the ion-acoustic {wave} behave like a constant
frequency wave for lower $\sigma$.}

\section{Nonlinear electrostatic waves}

{
When considering the formation of nonlinear electrostatic waves in such a plasma, we
assume that scale length of electrostatic structures is smaller than the average turbulence
scale length due to MRI, so that the effect of turbulence on these 
structures can be safely neglected. Thus, our plasma model consists of cold ions and
non-inertial electrons obeying Davydov velocity distribution. Though the negatively charged dust particles
is an integral component of the plasma in our model in the ion-acoustic time scale, dust dynamics 
can be neglected. However, their presence is quite necessary for the overall charge-balance.} 
The primary equation is the Poisson equation (dimensional),
\begin{equation}
\epsilon_{0}\frac{\partial^{2}\phi}{\partial x^{2}}=e(n_{e}-n_{i}+z_{d}n_{d}),\label{eq:Poisson}
\end{equation}
where $z_{d}$ is the dust-charge number and $n_{d}$ is the dust
density. The dust density remains constant, though the dust-charge
number $z_{d}$ must be determined self-consistently using the electron-ion
current balance equations to the dust particles. Following energy
conservation for the ions, the ion density $n_{i}$ can be written
as,
\begin{equation}
n_{i}=n_{i0}\left(1-\frac{2e\phi}{m_{i}u_{0}^{2}}\right)^{-1/2},\label{eq:iondensity}
\end{equation}
where we have assumed the ions to be cold and $u_{0}$ is the ion
velocity at $\infty$. The electron density is now given by Eq.(\ref{eq:totaldensity}).
As before, we choose to normalize the ion density by its equilibrium
values $n_{i}=n_{i0}$ , the plasma potential by $T_{e}/e$ and length
by Debye length $\lambda_{D}$. Far away from the perturbation, the
plasma potential vanishes, and we define the boundary conditions as
$\phi\to0$ at $x\to\infty$ and the normalized Poisson equation becomes,

\begin{equation}
\frac{\partial^{2}\phi}{\partial x^{2}}=n_{e}-\delta_{i}n_{i}+\delta_{d}z_{d}=F(\phi),\label{eq:normpoisson}
\end{equation}
where $\delta_{i}=n_{i0}/n_{e0}$ and $\delta_{d}=n_{d0}z_{d0}/n_{e0}$.
The dust-charge number $z_{d}$ is scaled as $z_{d}\to z_{d}/z_{d0}$,
where $z_{d0}$ is the equilibrium value of $z_{d}|_{\phi=0}$. Multiplying
the above equation by $d\phi/dx$ and integrating we have,
\begin{equation}
\frac{1}{2}\left(\frac{d\phi}{dx}\right)^{2}=\int_{0}^{\phi} {F(\psi)\,d\psi} =-V(\phi),\label{eq:sagdeev}
\end{equation}
 where $V(\phi)$ in Eq.(\ref{eq:sagdeev}) represents Sagdeev or
the pseudo-potential \citep{sagdeev}. The limits of the above integration
ensure that Sagdeev potential obeys the boundary condition at $\phi=0$,
namely $V(\phi=0)=0$. 

In the case of pure Maxwellian electrons, $n_{e}=e^{\phi}$ and the
integrations for electron and ion contributions in the above equation
can be carried out analytically and we have an expression for Sagdeev
potential as,
\begin{equation}
V(\phi)=1-e^{\phi}+\delta_{i}M^{2}\left(1-\sqrt{1-\frac{2\phi}{M^{2}}}\right)-\delta_{d}I_{d}(\phi),
\end{equation}
where $M=u_{0}/\sqrt{T_{e}/m_{i}}$ is the ion Mach number and
\begin{equation}
I_{d}(\phi)=\int_{0}^{\phi}z_{d}(\psi)\,d\psi\label{eq:dust}
\end{equation}
is the contribution of the dust particles to the pseudo-potential.
We note that the integral in Eq.(\ref{eq:sagdeev}) can not be carried
out analytically and has to be evaluated numerically. 

\textcolor{black}{A clarification on the definition of Mach number
}\textcolor{black}{\emph{must}}\textcolor{black}{{} be given at this point.
The Mach number here is defined in the case of usual ion-acoustic
speed for an Maxwellian electron-ion plasma. It has been argued in
the literature that the Mach number needs a re-definition in terms
of the acoustic speed,  defined in terms of the linear ion-acoustic
dispersion relation as
\begin{equation}
M=u_{0}/c_{{\rm dis}},
\end{equation}
where $c_{{\rm dis}}$ is the ion-acoustic speed determined from the
linear dispersion relation
\begin{equation}
c_{{\rm dis}}=\lim_{k\to0}\frac{\omega}{k}\equiv\lim_{k\to0}\frac{d\omega}{dk}.
\end{equation}
In this work, however, we shall continue to define the Mach number
with its so-called classical definition and will be careful about comparing
the soliton velocity }\textcolor{black}{\emph{only}}\textcolor{black}{{}
with the classical ion-acoustic speed.}

\subsection{Dust contribution}

In order to evaluate the integral given in Eq.(\ref{eq:dust}), we
assume that the net current (electron and ion) to the dust particles
remains zero
\begin{equation}
\sum_{j=i,e}I_{j}=0,\label{eq:total_current}
\end{equation}
where $I_{i,e}$ are the ion and electron currents to the surface
of the dust particles. \textcolor{black}{We note that the widely used
theoretical framework for calculation of dust-charging currents is
due to the orbit motion limited (OML) theory \citep{mott,allen}.
We further note that though the general basis for OML theory is not
affected by velocity distribution functions obeyed by the electrons
ions, the actual expression for currents will get changed by the use
of different velocity distribution functions. As such, in our case,
while the ion dust-charging current remains same as that for Maxwellian
ions, the electron dust-charging current will definitely be different.}

\begin{figure}
\begin{centering}
\includegraphics[clip,width=0.5\textwidth]{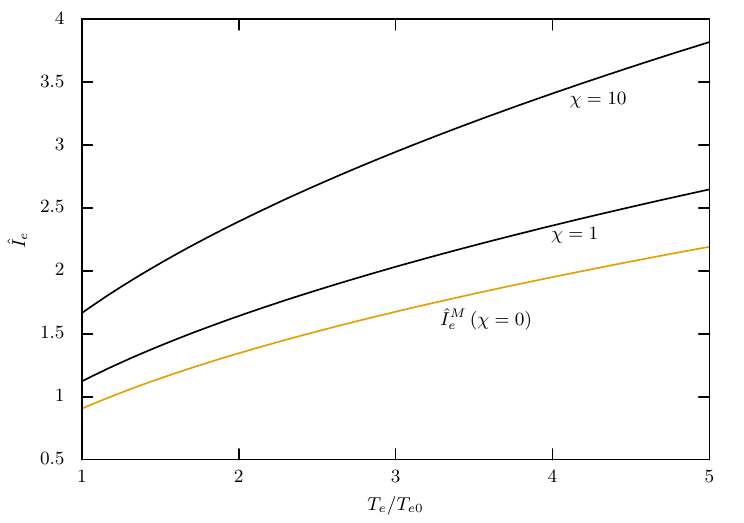}
\par\end{centering}
\caption{\protect\label{fig:Electron-dust-charging-current}Electron dust-charging
current for electrons with Davydov distribution.}
\end{figure}

\textcolor{black}{Following the orbit motion limited (OML) theory \citep{mott,allen,shukla},
the expressions for $I_{j}$ for a species `$j$' can be written as
(dimensional)
\begin{equation}
I_{j}=q_{j}\int_{v_{j}^{{\rm min}}}^{\infty}v_{j}\Sigma_{j}f_{j}(v_{j})\,d\bm{v}_{j},
\end{equation}
where $f_{j}(v_{j})$ is the velocity distribution function of the
corresponding charged particle and
\begin{equation}
\Sigma_{j}=\pi r_{d}^{2}\left(1-\frac{2q_{j}\varphi_{d}}{m_{j}v_{j}^{2}}\right)
\end{equation}
is the collision cross-section between a charged particle of charge
$q_{j}$ and a dust particle, where $r_{d}$ is the radius of a dust-particle
and $\varphi_{d}=\phi_{g}-\phi$ is the potential of the dust particles
with respect to the bulk plasma and $\phi_{g}$ being the grain potential.
The  minimum velocity with which the charged particle approaches
a dust particle is given by
\begin{equation}
v_{j}^{{\rm min}}=\left(\frac{2q_{j}\varphi_{d}}{m_{j}}\right)^{1/2}.
\end{equation}
So, for electrons with Davydov distribution function, we have
\begin{equation}
I_{e}=-4\pi qr_{d}^{2}n_{e}\int_{v_{e}^{{\rm min}}}^{\infty}\sigma v^{3}f_{e0}\,dv,\label{eq:Ied}
\end{equation}
where $f_{e0}$ is given by Eq.(\ref{eq:maxwellchi-1}). The above
integration can be evaluated analytically and can be written in terms
of upper incomplete gamma function $\Gamma(s,x)$
\begin{equation}
I_{e}=-4\pi r_{d}^{2}en_{e}\left(\frac{T_{e}}{2\pi m_{e}}\right)^{1/2}{\cal F}(e\varphi_{d}/T),
\end{equation}
where
\begin{equation}
{\cal {\cal F}}(x)=\frac{(\chi+x)\Gamma(\chi+1,\chi+x)-\Gamma(\chi+2,\chi+x)}{e^{-\chi}U\left(-\chi,-{\displaystyle \frac{1}{2}}-\chi,\chi\right)}.
\end{equation}
The ion dust-charging current retains its value for Maxwellian distribution
function \citep{shukla}
\begin{equation}
I_{i}=4\pi r_{d}^{2}en_{i}\left(\frac{T_{i}}{2\pi m_{i}}\right)^{1/2}\left(1-\frac{e\varphi_{d}}{T_{i}}\right),\label{eq:ii}
\end{equation}
In Fig.\ref{fig:Electron-dust-charging-current}, we have shown the
plots of $I_{e}$ as given by Eq.(\ref{eq:Ied}) with $T_{e}$ for
different $\chi$ and the equivalent expression for Maxwellian electrons
$I_{e}^{M}$
\begin{equation}
I_{e}^{M}=-4\pi r_{d}^{2}en_{e}\left(\frac{T_{e}}{2\pi m_{e}}\right)^{1/2}\exp\left(-\frac{e\varphi_{d}}{T_{e}}\right).\label{eq:ie}
\end{equation}
The resultant numerical results indicate that for small $\varphi_{d}$,
to a large extent, both the currents differ only by a multiplicative
factor and we can express
\begin{equation}
I_{e}\simeq\beta I_{e}^{M},
\end{equation}
where $\beta$ is a constant factor .}

Assuming the dust particles to be spherical in size, the net dust-charge
$q_{d}=-ez_{d}$ can be expressed in terms of dust potential $\varphi_{d}$
as
\begin{equation}
q_{d}=C\,\Delta V=4\pi\epsilon_{0}r_{d}\varphi_{d},
\end{equation}
where $C$ is the grain capacitance and. As per our definition
\begin{equation}
z_{d0}=4\pi\epsilon_{0}r_{d}e^{-1}\varphi_{d0}\label{eq:fid0}
\end{equation}
is the equilibrium dust-charge number at $\phi=0$, where $e$ is
the magnitude of electronic charge. The normalized expression for
dust potential is given by
\begin{equation}
\varphi_{d}=-\alpha^{-1}z_{d},
\end{equation}
where the dust potential is normalized by the magnitude of its equilibrium
value $\varphi_{d0}$, defined above and $\alpha$ is the magnitude
of the inverse of the normalized equilibrium dust potential $\alpha=|\varphi_{d0}|^{-1}$.
In normalised form, Eq.(\ref{eq:total_current}) can be written as\textcolor{black}{
\begin{equation}
\delta_{i}\delta_{m}\sigma^{1/2}n_{i}\left(1-\frac{\varphi_{d}}{\sigma}\right)-\beta n_{e}e^{\varphi_{d}}=0.\label{eq:charging}
\end{equation}
}where $\delta_{m}=\sqrt{m_{e}/m_{i}}\approx0.023$ and {$\sigma=T_{i0}/T_{e0}$
is the ratio of the equilibrium ion temperature to that of the electrons
and $\delta_i=n_{i0}/n_{e0}$ is the equilibrium ion to electron density ratio.} 
Using
the expression for ion density from Eq.(\ref{eq:iondensity}), one
can solve Eq.(\ref{eq:charging}) for $\varphi_{d}$ as a function
of $\phi$,\textcolor{black}{
\begin{equation}
\varphi_{d}(\phi)\simeq\sigma-W\left[\frac{\beta\sigma^{1/2}}{\delta_{i}\delta_{m}}e^{\sigma+\phi}\left(1-\frac{2\phi}{M^{2}}\right)^{1/2}\right],\label{eq:phid}
\end{equation}
}where $W(z)\equiv W_{0}(z)$ is the principal branch of Lambert $W$
function. One now needs to solve the integration in Eq.(\ref{eq:dust})
in order to find out the dust contribution to Sagdeev potential, which
however \emph{must} be evaluated numerically only. \textcolor{black}{The
full charging equation without the involvement of the multiplicative
factor $\beta$, can however be written in terms of the function ${\cal F}(x)$
described above for Davydov electrons as
\begin{equation}
\delta_{i}\delta_{m}\sigma^{1/2}n_{i}\left(1-\frac{\varphi_{d}}{\sigma}\right)-n_{e}{\cal F}(-\varphi_{d})=0.
\end{equation}
This equation however }\textcolor{black}{\emph{must}}\textcolor{black}{{}
be solved numerically for $\varphi_{d}$. In what follows, we shall
approximate the dust potential from Eq.(\ref{eq:phid}), which helps
us tackle the problem analytically.}

The pseudo-potential has to be evaluated numerically from the following
differential equation with the initial condition $V(0)=0$,
\begin{equation}
\frac{d}{d\phi}V(\phi)=-F(\phi),\quad V(0)=0,
\end{equation}
where
\begin{eqnarray}
F(\phi)&=&e^{\phi}\frac{U\left(-\chi,-{\displaystyle \frac{1}{2}}-\chi,-\phi+\chi\right)}{U\left(-\chi,-{\displaystyle \frac{1}{2}}-\chi,\chi\right)}-\delta_{i}\left(1-\frac{2\phi}{M^{2}}\right)^{-1/2}\nonumber\\
&&-\,\alpha\delta_{d}\left(\sigma-W\left[{\cal Z}e^{\phi}\left(1-\frac{2\phi}{M^{2}}\right)\right]\right),
\end{eqnarray}
where $\alpha=|\varphi_{d0}|^{-1}$ is given by\textcolor{black}{
\begin{eqnarray}
\alpha & = & \left|\sigma-W({\cal Z})\right|^{-1},\\
{\cal Z} & = & e^{\sigma}\frac{\beta\sigma^{1/2}}{\delta_{i}\delta_{m}}
\end{eqnarray}
}

\begin{figure}
\begin{centering}
\includegraphics[clip,width=0.5\textwidth]{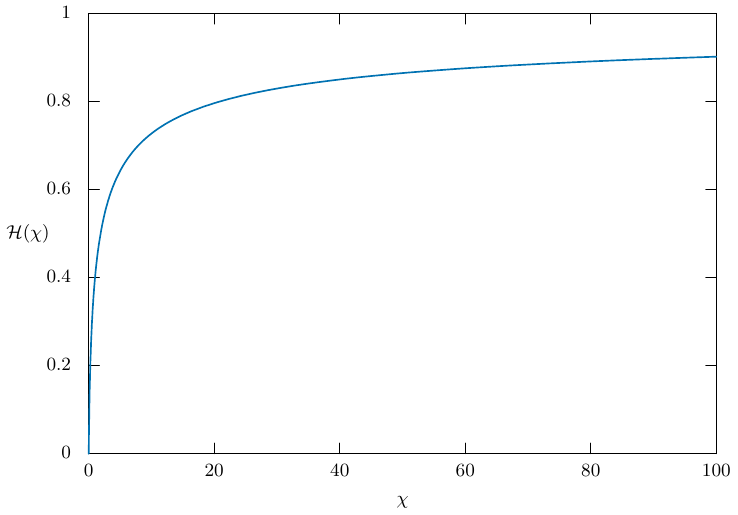}\hfill{}\includegraphics[clip,width=0.5\textwidth]{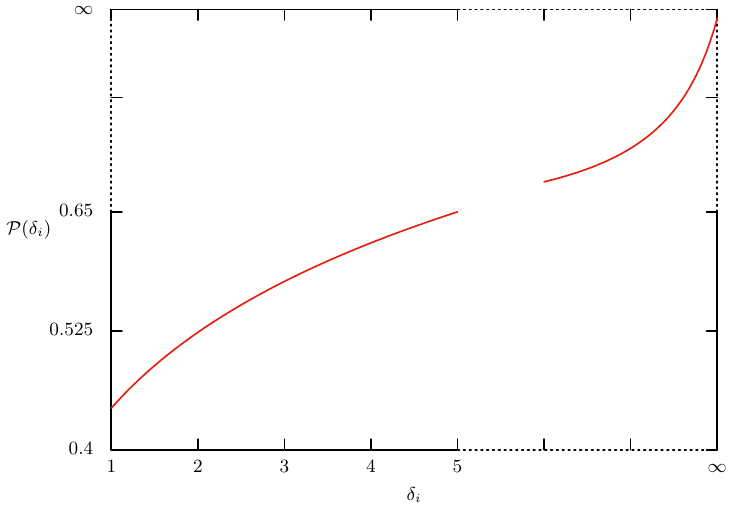}
\par\end{centering}
\caption{\protect\label{fig:The-functions-}The functions ${\cal H}$ (left)
and ${\cal P}$ (right for $\beta=1$). }
\end{figure}

\section{Solitary waves with dust effects}

\subsection{The Maxwellian limit $(\chi=0)$ without dust}

This is the classical limit for an $e$-$i$ plasma. In this limit
$\chi=0$ and the function $F(\phi)$ reduces to
\begin{equation}
F(\phi)=e^{\phi}-\left(1-\frac{2\phi}{M^{2}}\right)^{-1/2}
\end{equation}
and the pseudo-potential can be found analytically as
\begin{equation}
V_{\textrm{Max}}(\phi)=1-e^{\phi}+M^{2}\left(1-\sqrt{1-\frac{2\phi}{M^{2}}}\right).\label{eq:vmax}
\end{equation}
{By expanding the above expression around $\phi=0$, we can show that $V_{\rm Max}>0$ as $\phi\to\pm0$ for $M<1$. For $M>1$, however $V_{\rm Max}$ always monotonically decreases for $\phi<0$, so that we can conclude that 
 dark soliton $(\phi<0)$ cannot
exist.  At the same time, we can also see that for $V_{\rm Max}$ to have a potential structure for $\phi>0$, $M$ must be $>1$. For $(\phi>0)$, we can readily establish the lower and upper
limits for Mach number $M$}
\begin{equation}
1<M\leq1.5852.
\end{equation}
{ The upper value comes from the fact that for $M\simeq1.5852$, $V_{\rm Max}(\phi)$ becomes exactly vertical (slope $\to\infty$)  when simultaneously $V_{\rm Max}(\phi)=0$. The value can be found out by equating $\phi\equiv\phi_m=M^2/2$ (the condition for $dV_{\rm Max}/d\phi\to\infty$) in Eq.(\ref{eq:vmax}) and solve for $M$. The value of $\phi_m$ can be found by substituting $M=\sqrt{2\phi_m}$ in Eq.(\ref{eq:vmax}), which can be solved analytically}
\begin{equation}
\phi_{m}={\displaystyle -\frac{1}{2}-W_{-1}\left(-\frac{1}{2\sqrt{e}}\right)}\simeq1.2564,
\end{equation}
with $e$ being the Euler's number and $W_{-1}(z)$ being the negative
branch of Lambert $W$ function. 

{The existence of bright solitons $(\phi>0)$ is shown in Fig.\ref{fig:The-pseudo-potential-curves} in terms of pseudo-potential in Section 6.3, where we have numerically reconstructed the same.}

\subsection{With Davydov electrons}

In order to have an idea of the limits of Mach numbers for Davydov
electrons with dust effects, we look at the behavior of the second
order derivative of the pseudo-potential at $\phi=0$, which is given
by,
\begin{equation}
V''(0)\equiv\left.\frac{d^{2}V}{d\phi^{2}}\right|_{\phi=0}=-1+\frac{\delta_{i}}{M^{2}}+{\cal H}(\chi)-\delta_{d}\left(1-\frac{1}{M^{2}}\right){\cal P}(\delta_{i})
\end{equation}
where\textcolor{black}{
\begin{eqnarray}
{\cal H}(\chi) & = & \chi\frac{U\left(1-\chi,{\displaystyle \frac{1}{2}}-\chi,\chi\right)}{U\left(-\chi,-{\displaystyle \frac{1}{2}}-\chi,\chi\right)},\\
{\cal P}(\delta_{i}) & = & \frac{W({\cal Z})}{|\sigma-W({\cal Z})|[1+W({\cal Z})]}.
\end{eqnarray}
}We note that the function ${\cal H}(\chi)$ is always bounded in
the region $\chi\in[0,+\infty)$ with $0\leq{\cal H}(\chi)<1$, whereas
the function ${\cal P}(\delta_{i})$ is a positive and monotonically
increasing function of $\delta_{i}$. The balance of these two functions
will provide the potential-well nature of the pseudo-potential. The
plots of both these functions are shown in Fig.\ref{fig:The-functions-}.
Demanding $V'(0)\leq0$, we get the lower limit for the Mach number
\begin{equation}
M>M_{-}=\left(\frac{\delta_{d}{\cal P}+\delta_{i}}{\delta_{d}{\cal P}+1-{\cal H}}\right)^{1/2}.\label{eq:ml}
\end{equation}
The upper limit on the Mach number $M_{+}$ is given by the expression
\begin{equation}
M_{+}=\sqrt{2\phi_{\textrm{max}}},
\end{equation}
where $\phi_{\textrm{max}}$ is a solution of the equation $V(\phi)=0|_{M^{2}\to2\phi}$.
Apparently, this equation in general, can be written as
\begin{eqnarray}
2\delta_{i}\phi-\int_{0}^{\phi}e^{\psi}\frac{U\left({-\chi},{\displaystyle \frac{1}{2}}-\chi,-\psi+\chi\right)}{U\left(-\chi,-{\displaystyle \frac{1}{2}}-\chi,\chi\right)}\,d\psi &&\nonumber\\
+\,\alpha\delta_{d}\int_{0}^{\phi}{\left\{\sigma-W\left[{\cal Z}e^\psi\left(1-\frac{\psi}{\phi}\right)^{1/2}\right]\right\}}\,d\psi&=&0.
\end{eqnarray}
However, for the existence of $M_{+}$ and thereby $\phi_{\textrm{max}}$,
the above equation \emph{must} have a real solution $\phi=\phi_{\textrm{max}}\neq0$,
such that $\chi\geq\phi$ in the domain $\phi\in[0,\phi_{\textrm{max}}]$.

\begin{figure}
\begin{centering}
\includegraphics[clip,width=0.5\textwidth]{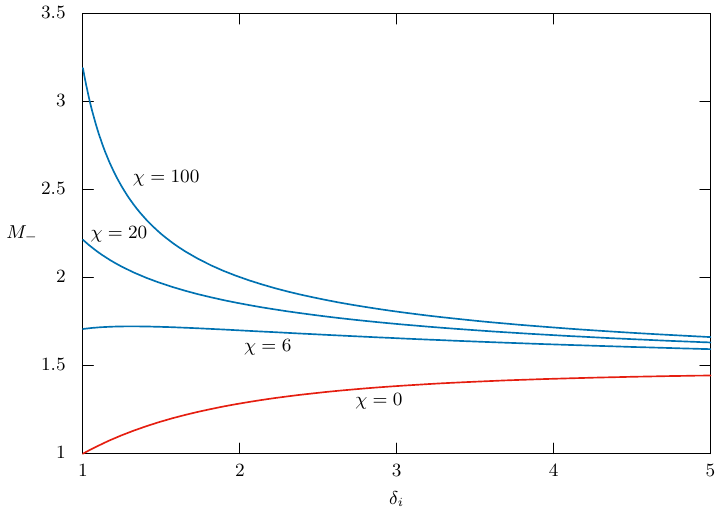}\hfill{}\includegraphics[clip,width=0.5\textwidth]{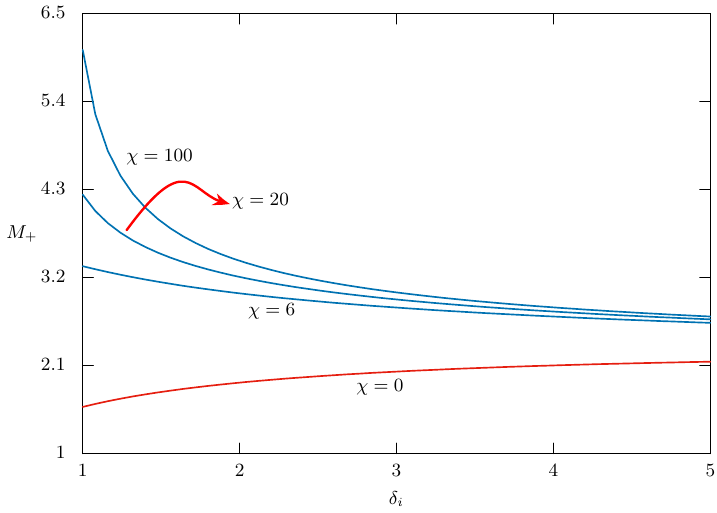}
\par\end{centering}
\caption{\protect\label{fig:The-lower-}The lower $(M_{-})$ and upper $(M_{+})$
limits for the Mach number as dust concentration increases for various
values of $\chi$. }
\end{figure}

\textcolor{black}{In the numerical results that we are going to show,
we use the constant factor $\beta=1$, as the important and significant
results are always in the region $\delta_{i}\sim1$, where $\beta\sim1$}.
In Fig.\ref{fig:The-lower-}, we have plotted the $M_{\pm}$ versus
$\delta_{i}$ for various $\chi$. The distinguishing feature for
non-Maxwellian electrons $(\chi>0)$ is that both the lower and upper
limits of the Mach number decrease with increasing $\delta_{i}$ while
for Maxwellian plasma the trend is opposite. We can also see from
the plots of $M_{-}$, that for $\chi\gtrsim10$, the Mach number
goes beyond 2 and can become more than 3 for higher $\chi$ for low
dust concentration. As we note that even at high values of $\delta_{i}\sim2$,
$M_{-}$ can stay above 2 for higher $\chi$. We can refer to these
solitons as \emph{hypersonic} \textcolor{black}{in comparison to a
classical Maxwellian $e$-$i$ plasma \citep{chen-1974}}, as the minimum velocity with
which these solitons travel is more than twice the ion-sound speed\emph{.}
We also note that for $\delta_{i}>1$, such \textcolor{black}{\emph{hypersonic}}
solitons can also have very large amplitudes. So, there is a possibility
of fast-moving large amplitude solitons, reaching large distances,
which will eventually dissipate and can give rise to the so-called
soliton radiation \citep{rad}. 

\begin{figure}
\begin{centering}
\includegraphics[clip,width=0.5\textwidth]{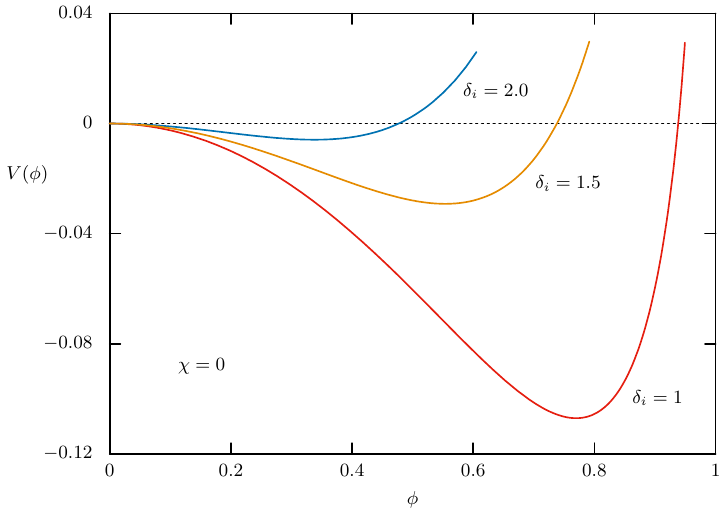}\hfill{}\includegraphics[clip,width=0.5\textwidth]{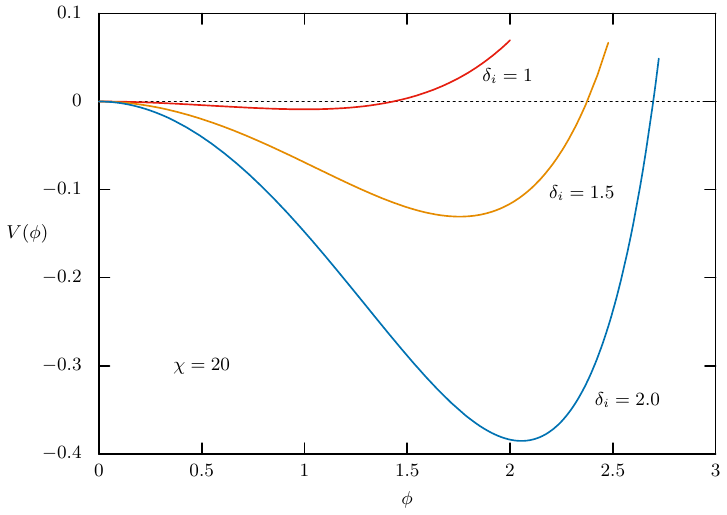}
\par\end{centering}
\caption{\protect\label{fig:The-pseudo-potential-curves}The pseudo-potential
curves for two cases -- Maxwellian ($\chi=0$, left) and Davydov
electrons ($\chi=20$, right) for different $\delta_{i}$. Note the
contrasting behavior of the soliton amplitudes with increasing $\delta_{i}$
in both cases.}
\end{figure}

In Fig.\ref{fig:The-pseudo-potential-curves}, we show the pseudo-potential
curves for two cases -- Maxwellian electrons $(\chi=0)$ and Davydov
electrons with $\chi=20$. The contrasting behavior of these two cases
as already shown in Fig.\ref{fig:The-lower-}, can be seen from the
variation of soliton amplitude with $\delta_{i}$. We note that the
soliton amplitude is given by the first zero of the pseudo-potential
away from $\phi=0$. While for Maxwellian electrons the soliton amplitude
decreases with increasing dust concentration, it increases for $\chi>0$.
Moreover, as $\chi$ increases, the solitons become large but slow
down considerably, which can be understood from the conservation of
energy. But it is important to note that for lower dust concentration
and higher $\chi$, we have $M>2$. So, high-energy packets of charge-perturbation
can travel very fast in this situation.

\subsection{Numerical reconstruction of pseudo-potential}

\begin{figure}
\begin{centering}
\includegraphics[width=0.5\textwidth]{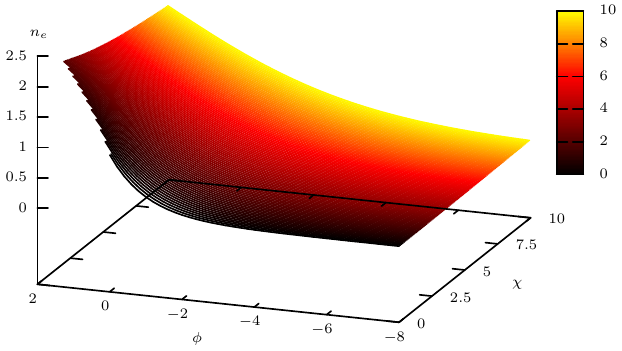}\hfill{}\includegraphics[width=0.5\textwidth]{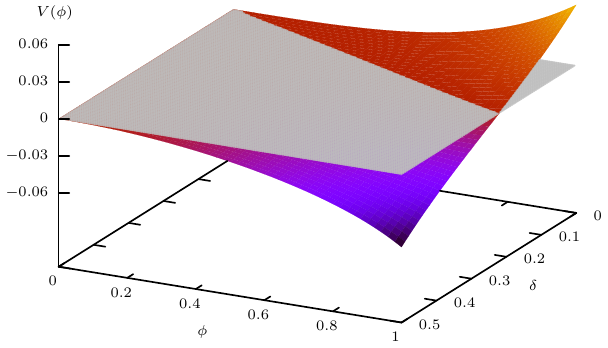}
\par\end{centering}
\caption{\protect\label{fig:A-mapping-in}Electron density $n_{e}$ as a function
of $(\chi,\phi)$. All the variables are in normalized units (left).
The intersection of the $V(\phi)=0$ plan with surface mapped by $(V,\delta,\phi)$
space for $\delta>0$ and $\phi>0$ (right).}
\end{figure}

As we have seen, for arbitrary $(\chi,\delta_{i})$, the pseudo-potential
has to be constructed numerically. We now outline a procedure, which
in general, can be utilized to construct the pseudo-potential with
complicated density expressions. This can then be used to solve Poisson's
equation. We shall do this through an \emph{inverse mapping} of the
electron density function $n_{e}\equiv F_{e}(\phi)$ given in Eq.(\ref{eq:normpoisson}).
To start with, we consider a mapping of the region in the $(\chi,\phi,n_{e})$
space for $\delta_{i}=1$. In Fig.\ref{fig:A-mapping-in}, we show
such a mapping, obtained numerically.

\begin{figure}
\begin{centering}
\includegraphics[width=0.5\textwidth]{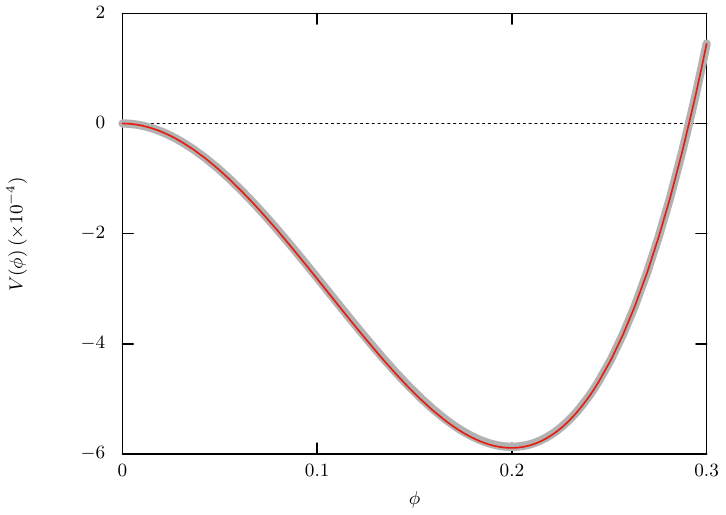}\hfill{}\includegraphics[width=0.5\textwidth]{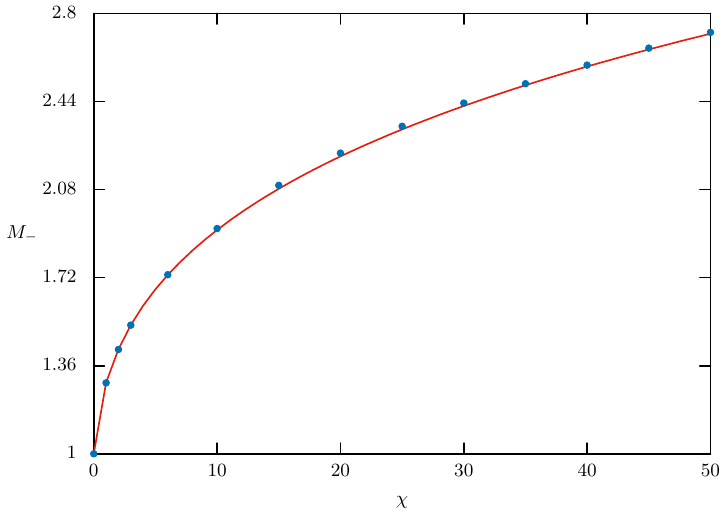}
\par\end{centering}
\caption{\protect\label{fig:Reconstructed-Sagdeev-potential}Left: Reconstructed
pseudo-potential through (red solid line) inverse-mapping for $\chi=1$,
superimposed over the analytically calculated one (thick gray line)
for $M=1.4$. Right: The reconstruction of $M_{-}$ as given by expression
(\ref{eq:mlow}) shown in circle, superimposed with the analytically
calculated $M_{-}$ from Eq.(\ref{eq:ml}).}
\end{figure}

It turns out that for any arbitrary $\chi$, the inverse function
$F_{e}^{-1}(n_{e})$ can be fitted into an inverse $\tanh$ function
as
\begin{equation}
\phi\equiv F_{e}^{-1}(n_{e})=\frac{1}{b}\left[\tanh^{-1}\left(\frac{n_{e}-a_{1}}{a_{2}}\right)-a_{3}\right],
\end{equation}
corresponding to map
\begin{equation}
n_{e}=a_{1}+a_{2}\,\tanh(b\phi+a_{3}),\label{eq:nfit}
\end{equation}
where $a_{i}$s are constants as determined through the nonlinear
fitting. The fitting itself is carried out with the help of the Mathematica
algebra system. However, we note that the asymptotic limits of $n_{e}$
at $\phi\to(-\infty,0]$ are $(0,1]$. Using these values, we see
that
\begin{equation}
a_{1}=a_{2}=a,\quad a_{3}=\tanh^{-1}\left(\frac{1}{a}-1\right),
\end{equation}
so that we have the inverse map as
\begin{equation}
\phi=\frac{1}{b}\left[{\tanh^{-1}\left(1-\frac{1}{a}\right)}+\tanh^{-1}\left(\frac{n_{e}}{a}-1\right)\right].\label{eq:phifit}
\end{equation}
This enables us to find out an analytical form for the Sagdeev potential,
\begin{eqnarray}
V(\phi)&\simeq &M^{2}\left[1-\left(1-\frac{2\phi}{M^{2}}\right)^{1/2}\right]-\frac{a}{2b}\log\left(\frac{2a-1}{a^{2}}\right)-a\phi\nonumber\\
& & -\,\frac{a}{b}\log\left[\cosh\left\{ b\phi+\tanh^{-1}\left(\frac{1}{a}-1\right)\right\} \right].\label{eq:vphi}
\end{eqnarray}
At the Maxwellian limit, $a\to\infty,b\to1/2$ and the expression
in Eq.(\ref{eq:nfit}), $n_{e}\to e^{\phi}$. It can also be seen
that for all other cases, we have 
\begin{equation}
\infty>a>1,\quad\frac{1}{2}>b>0.
\end{equation}
Computationally, a nonlinear fitting through Eq.(\ref{eq:phifit})
for $\chi=1$, yields $a\simeq1.44854,b=0.449851$ for which we have
shown the inverse-mapping-computed Sagdeev potential along with the
analytically calculated one in Fig.\ref{fig:Reconstructed-Sagdeev-potential}
(left) which shows an excellent correspondence between the two. {The global fitting error remains $\lesssim10^{-4}$.}

To find the limits on $M$, we carry out a series expansion for $V(\phi)$
as given by Eq.(\ref{eq:vphi}) near $\phi\sim0$, it can be shown
that a soliton can exist only if
\begin{equation}
M>M_{-}=\left(\frac{a/b}{2a-1}\right)^{1/2},\label{eq:mlow}
\end{equation}
which reduces to the Maxwellian condition $M_{-}^{M}=M>1$ at the
limits $a\to\infty,b\to1/2$. This provides us with the lower limit
for $M$. The limit as obtained above is compared to the actual analytical
expression given in Eq.(\ref{eq:ml}) in Fig.\ref{fig:Reconstructed-Sagdeev-potential}
(right). 

The derivative of the reconstructed Sagdeev potential with respect
to $\phi$ is given by

\begin{equation}
\frac{d}{d\phi}V(\phi)\equiv V'(\phi)=\left(1-\frac{2\phi}{M^{2}}\right)^{-1/2}-a\left[1+\tanh\left\{ b\phi+\tanh^{-1}\left(\frac{1}{a}-1\right)\right\} \right],
\end{equation}
From the conditions imposed on $(a,b)$, we can see that for $\phi<0$,
\begin{eqnarray}
b\phi+\tanh^{-1}\left(\frac{1}{a}-1\right) & < & 0,\\
0\leq1+\tanh\left\{ b\phi+\tanh^{-1}\left(\frac{1}{a}-1\right)\right\}  & \leq & 1,\\
\left(1-\frac{2\phi}{M^{2}}\right)^{-1/2} & \leq & 1,
\end{eqnarray}
from which, it can be numerically shown that for the intervals $a\in[1,\infty)$
and $b\in(0,1/2]$
\begin{equation}
V'(\phi)>0,\quad\forall(a,b),\phi<0,
\end{equation}
so that $V(\phi)$ \emph{can not} have any local (or global) maxima
for $\phi<0$ and we conclude that for all values of $\chi$, dark
solitons cannot exist.

To see, if we have any upper limit for $M$, we expand $V(\phi)$
about $M=M_{-}$
\begin{eqnarray}
V(\phi) & \simeq & M_{-}^{2}(1-\alpha)-a\phi-\frac{a}{2b}\ln\left[\left(\frac{2a-1}{a^{2}}\right)\cosh^{2}\left\{ b\phi+\tanh^{-1}\left(\frac{1}{a}-1\right)\right\} \right]\nonumber \\
 &  & +\,2M_{-}\left[1-\alpha\left(1+\frac{\phi}{M_{-}^{2}-2\phi}\right)\right]\delta+\left[1-\alpha M_{-}^{2}\frac{\left(M_{-}^{2}-3\phi\right)}{\left(M_{-}^{2}-2\phi\right)^{2}}\right]\delta^{2}\nonumber\\
 &  & -\,\delta^{3}\frac{2\alpha M_{-}\phi^{2}}{\left(M_{-}^{2}-2\phi\right)^{3}}+{\cal O}(\delta^{4}),
\end{eqnarray}
where $\delta=M-M_{-}>0$ is the deviation of $M$ from $M_{-}$.
As $V(\phi)<0$ in the neighborhood of $\phi\sim0$ for $M>M_{-}$,
it must be again positive for a certain value of $\phi=\phi_{m}$
away from $\phi=0$ for a soliton to exist. It can now be numerically
shown that the surface mapped by $(V,\delta,\phi)$ space for $\delta>0$
and $\phi>0$ intersects with the $V(\phi)=0$ plan only when $\delta\to\delta_{\textrm{max}}$,
as shown in Fig.\ref{fig:A-mapping-in} (right) This shows that there
exists an upper value for $M=M_{+}\simeq M_{-}+\delta_{\textrm{max}}$
for the existence of solitons.

\section{Conclusion}

To summarize, we have considered the possibility of the formation
of electrostatic solitons in a plasma environment, which is characterized
by MRI-driven turbulent heating of the electrons, making the electrons
obey a non-Maxwellian velocity distribution, particularly the so-called
Davydov distribution. The electron distribution becomes Maxwellian
when the heating is neglected and becomes a Druyvesteyn distribution
function \citep{Druyvesteyn-1940} in the extreme case. The plasma
itself is described by such electrons, a Maxwellian ion fluid and
negatively charged dust grains. The dust grains are assumed to be
static in the ion-acoustic timescale and ion temperature is neglected
in comparison to that of the electrons. We show that the situation
for such a condition is favorable in the plasma environment of a PPD,
specifically in the MRI-stable dead zones, which are rife with turbulence-driven
random heating. We show that electrostatic nonlinear structures, in
these situations, can form with very large velocities, often crossing
several orders of magnitude larger than the ion-sound speed. These
fast-moving solitons can also be of quite high amplitudes. We argue
that these  solitons can be an effective mechanism for energy equilibration
through a process known as soliton radiation. One of the interesting
findings of this analysis is that the behavior of soliton velocity
(the Mach number) with dust density reverses as the electron velocity
distribution deviates more from Maxwellian. We have however neglected
the dust-charge fluctuation and assumed that the net current to the
dust particles is always zero. It is to be also noted that $\chi$,
which is a measure of how high is the heating, can be easily estimated
to reach up to about $\sim100$ \citep{Okuzumi-2015}, making the
electron highly non-Maxwellian. With relevant parameters for PPDs,
we find the soliton velocity in extreme cases can become more than
$\sim4-5$ times the ion-sound speed and their amplitudes can reach
up to several hundreds of Debye length.

We have also outlined a procedure to facilitate analytical analysis
of the resultant pseudo-potential through an inverse-mapping of the
electron density with plasma potential. This procedure makes the analysis
easier where an analytic treatment of the densities in terms of plasma
potential is not otherwise feasible due to their complicated expressions,
such as one encountered in this work.

\section*{Acknowledgement}
It is a pleasure to thank the anonymous referees for their suggestions, which have greatly
improved the manuscript.

\section*{Declarations}

{\bf Author Contributions:} All authors contributed equally.\\

\noindent
{\bf Data Availability Statement:} No Data associated in the manuscript.\\

\noindent
{\bf Conflicts of Interest:} The authors declare no conflict of interest.


\begin{thebibliography}{52}
\ifx \bisbn   \undefined \def \bisbn  #1{ISBN #1}\fi
\ifx \binits  \undefined \def \binits#1{#1}\fi
\ifx \bauthor  \undefined \def \bauthor#1{#1}\fi
\ifx \batitle  \undefined \def \batitle#1{#1}\fi
\ifx \bjtitle  \undefined \def \bjtitle#1{#1}\fi
\ifx \bvolume  \undefined \def \bvolume#1{\textbf{#1}}\fi
\ifx \byear  \undefined \def \byear#1{#1}\fi
\ifx \bissue  \undefined \def \bissue#1{#1}\fi
\ifx \bfpage  \undefined \def \bfpage#1{#1}\fi
\ifx \blpage  \undefined \def \blpage #1{#1}\fi
\ifx \burl  \undefined \def \burl#1{\textsf{#1}}\fi
\ifx \doiurl  \undefined \def \doiurl#1{\url{https://doi.org/#1}}\fi
\ifx \betal  \undefined \def \betal{\textit{et al.}}\fi
\ifx \binstitute  \undefined \def \binstitute#1{#1}\fi
\ifx \binstitutionaled  \undefined \def \binstitutionaled#1{#1}\fi
\ifx \bctitle  \undefined \def \bctitle#1{#1}\fi
\ifx \beditor  \undefined \def \beditor#1{#1}\fi
\ifx \bpublisher  \undefined \def \bpublisher#1{#1}\fi
\ifx \bbtitle  \undefined \def \bbtitle#1{#1}\fi
\ifx \bedition  \undefined \def \bedition#1{#1}\fi
\ifx \bseriesno  \undefined \def \bseriesno#1{#1}\fi
\ifx \blocation  \undefined \def \blocation#1{#1}\fi
\ifx \bsertitle  \undefined \def \bsertitle#1{#1}\fi
\ifx \bsnm \undefined \def \bsnm#1{#1}\fi
\ifx \bsuffix \undefined \def \bsuffix#1{#1}\fi
\ifx \bparticle \undefined \def \bparticle#1{#1}\fi
\ifx \barticle \undefined \def \barticle#1{#1}\fi
\bibcommenthead
\ifx \bconfdate \undefined \def \bconfdate #1{#1}\fi
\ifx \botherref \undefined \def \botherref #1{#1}\fi
\ifx \url \undefined \def \url#1{\textsf{#1}}\fi
\ifx \bchapter \undefined \def \bchapter#1{#1}\fi
\ifx \bbook \undefined \def \bbook#1{#1}\fi
\ifx \bcomment \undefined \def \bcomment#1{#1}\fi
\ifx \oauthor \undefined \def \oauthor#1{#1}\fi
\ifx \citeauthoryear \undefined \def \citeauthoryear#1{#1}\fi
\ifx \endbibitem  \undefined \def \endbibitem {}\fi
\ifx \bconflocation  \undefined \def \bconflocation#1{#1}\fi
\ifx \arxivurl  \undefined \def \arxivurl#1{\textsf{#1}}\fi
\csname PreBibitemsHook\endcsname

\bibitem[\protect\citeauthoryear{Pringle}{1981}]{pringle}
\begin{barticle}
\bauthor{\bsnm{Pringle}, \binits{J.E.}}
\bjtitle{Ann. Rev. Astron. Astrophys.}
\bvolume{19},
\bfpage{137}
(\byear{1981})
\end{barticle}
\endbibitem

\bibitem[\protect\citeauthoryear{Mamajek et~al.}{2004}]{eric}
\begin{barticle}
\bauthor{\bsnm{Mamajek}, \binits{E.E.}},
\bauthor{\bsnm{Meyer}, \binits{M.R.}},
\bauthor{\bsnm{Hinz}, \binits{P.M.}},
\bauthor{\bsnm{Hoffmann}, \binits{W.F.}},
\bauthor{\bsnm{Cohen}, \binits{M.}},
\bauthor{\bsnm{Hora}, \binits{J.L.}}
\bjtitle{Astrophys. J.}
\bvolume{612},
\bfpage{496}
(\byear{2004})
\end{barticle}
\endbibitem

\bibitem[\protect\citeauthoryear{White and Hillenbrand}{2005}]{russel}
\begin{barticle}
\bauthor{\bsnm{White}, \binits{R.J.}},
\bauthor{\bsnm{Hillenbrand}, \binits{L.A.}}
\bjtitle{Astrophys. J.}
\bvolume{621},
\bfpage{65}
(\byear{2005})
\end{barticle}
\endbibitem

\bibitem[\protect\citeauthoryear{Armitage}{2011}]{armitage}
\begin{barticle}
\bauthor{\bsnm{Armitage}, \binits{P.J.}}
\bjtitle{Ann. Rev. Astron. Astrophys.}
\bvolume{49},
\bfpage{195}
(\byear{2011})
\end{barticle}
\endbibitem

\bibitem[\protect\citeauthoryear{Lesur}{2021}]{lesur}
\begin{barticle}
\bauthor{\bsnm{Lesur}, \binits{G.R.J.}}
\bjtitle{J. Plasma Phys.}
\bvolume{87},
\bfpage{205870101}
(\byear{2021})
\end{barticle}
\endbibitem

\bibitem[\protect\citeauthoryear{Muranushi}{2010}]{taka}
\begin{barticle}
\bauthor{\bsnm{Muranushi}, \binits{T.}}
\bjtitle{Mon. Not. R. Astron. Soc.}
\bvolume{401},
\bfpage{2641}
(\byear{2010})
\end{barticle}
\endbibitem

\bibitem[\protect\citeauthoryear{Akimkin et~al.}{2023}]{vitaly}
\begin{barticle}
\bauthor{\bsnm{Akimkin}, \binits{V.}},
\bauthor{\bsnm{Ivlev}, \binits{A.V.}},
\bauthor{\bsnm{Caselli}, \binits{P.}},
\bauthor{\bsnm{Gong}, \binits{M.}},
\bauthor{\bsnm{Silbee}, \binits{K.}}
\bjtitle{Astrophys. J.}
\bvolume{953},
\bfpage{72}
(\byear{2023})
\end{barticle}
\endbibitem

\bibitem[\protect\citeauthoryear{Balbus and Hawley}{1991}]{Balbus-1991}
\begin{botherref}
\oauthor{\bsnm{Balbus}, \binits{S.A.}},
\oauthor{\bsnm{Hawley}, \binits{J.F.}}
ApJ
\textbf{376}(214)
(1991)
\end{botherref}
\endbibitem

\bibitem[\protect\citeauthoryear{Hawley et~al.}{1995}]{Hawley-1995}
\begin{botherref}
\oauthor{\bsnm{Hawley}, \binits{J.F.}},
\oauthor{\bsnm{Gammie}, \binits{C.F.}},
\oauthor{\bsnm{Balbus}, \binits{S.A.}}
ApJ
\textbf{440}(742)
(1995)
\end{botherref}
\endbibitem

\bibitem[\protect\citeauthoryear{Gammie}{1996}]{gammie}
\begin{barticle}
\bauthor{\bsnm{Gammie}, \binits{F.C.}}
\bjtitle{Astrophys. J.}
\bvolume{457},
\bfpage{355}
(\byear{1996})
\end{barticle}
\endbibitem

\bibitem[\protect\citeauthoryear{Wardle}{2007}]{wardle}
\begin{barticle}
\bauthor{\bsnm{Wardle}, \binits{M.}}
\bjtitle{Astrophys. J. Suppl.}
\bvolume{311},
\bfpage{35}
(\byear{2007})
\end{barticle}
\endbibitem

\bibitem[\protect\citeauthoryear{Inutsuka and Sano}{2005}]{Inutsuka-2005}
\begin{botherref}
\oauthor{\bsnm{Inutsuka}, \binits{S.-I.}},
\oauthor{\bsnm{Sano}, \binits{T.}}
ApJL
\textbf{628}(L155)
(2005)
\end{botherref}
\endbibitem

\bibitem[\protect\citeauthoryear{Okuzumi and Inutsuka}{2015}]{Okuzumi-2015}
\begin{botherref}
\oauthor{\bsnm{Okuzumi}, \binits{S.}},
\oauthor{\bsnm{Inutsuka}, \binits{S.-I.I.}}
Astrophys. J.
\textbf{800}(47)
(2015)
\end{botherref}
\endbibitem

\bibitem[\protect\citeauthoryear{Golant et~al.}{1980}]{golant}
\begin{bbook}
\bauthor{\bsnm{Golant}, \binits{V.E.}},
\bauthor{\bsnm{Zhilinsky}, \binits{A.P.}},
\bauthor{\bsnm{Sakharov}, \binits{I.E.}}:
\bbtitle{Fundamentals of Plasma Physics}.
\bpublisher{Wiley},
\blocation{New York}
(\byear{1980})
\end{bbook}
\endbibitem

\bibitem[\protect\citeauthoryear{Lifshitz and Pitaevskii}{1981}]{Lifshitz-1981}
\begin{bbook}
\bauthor{\bsnm{Lifshitz}, \binits{E.M.}},
\bauthor{\bsnm{Pitaevskii}, \binits{L.P.}}:
\bbtitle{Physical Kinetics}
vol. \bseriesno{10}.
\bpublisher{Pergamon Press},
\blocation{Oxford}
(\byear{1981})
\end{bbook}
\endbibitem

\bibitem[\protect\citeauthoryear{Davydov}{1935}]{Davydov-1935}
\begin{botherref}
\oauthor{\bsnm{Davydov}, \binits{B.}}
Phys. Z. Sowjet.
\textbf{8}(59)
(1935)
\end{botherref}
\endbibitem

\bibitem[\protect\citeauthoryear{Galeev and Sagdeev}{1979}]{galeev}
\begin{barticle}
\bauthor{\bsnm{Galeev}, \binits{A.A.}},
\bauthor{\bsnm{Sagdeev}, \binits{R.Z.}}
\bjtitle{Rev. Plasma Phys.}
\bvolume{7},
\bfpage{1}
(\byear{1979})
\end{barticle}
\endbibitem

\bibitem[\protect\citeauthoryear{Kuzentsov et~al.}{1986}]{stability}
\begin{barticle}
\bauthor{\bsnm{Kuzentsov}, \binits{E.A.}},
\bauthor{\bsnm{Rubenchik}, \binits{A.M.}},
\bauthor{\bsnm{Zakharov}, \binits{V.E.}}
\bjtitle{Phys. Rep.}
\bvolume{142}(\bissue{3}),
\bfpage{103}
(\byear{1986})
\end{barticle}
\endbibitem

\bibitem[\protect\citeauthoryear{Karpman and Schamel}{1997}]{schamel}
\begin{barticle}
\bauthor{\bsnm{Karpman}, \binits{V.I.}},
\bauthor{\bsnm{Schamel}, \binits{H.}}
\bjtitle{Phys. Plasmas}
\bvolume{4},
\bfpage{120}
(\byear{1997})
\end{barticle}
\endbibitem

\bibitem[\protect\citeauthoryear{Murusidze et~al.}{1998}]{rad}
\begin{barticle}
\bauthor{\bsnm{Murusidze}, \binits{I.G.}},
\bauthor{\bsnm{Tsintsadze}, \binits{N.L.}},
\bauthor{\bsnm{Tskhakaya}, \binits{D.D.}},
\bauthor{\bsnm{Shukla}, \binits{P.K.}}
\bjtitle{Phys. Scr.}
\bvolume{58},
\bfpage{266}
(\byear{1998})
\end{barticle}
\endbibitem

\bibitem[\protect\citeauthoryear{Antonova}{2008}]{turb}
\begin{barticle}
\bauthor{\bsnm{Antonova}, \binits{E.E.}}
\bjtitle{Adv. Space Res.}
\bvolume{41}(\bissue{10}),
\bfpage{1677}
(\byear{2008})
\end{barticle}
\endbibitem

\bibitem[\protect\citeauthoryear{Rubinstein and Clark}{2017}]{rubinstein}
\begin{barticle}
\bauthor{\bsnm{Rubinstein}, \binits{R.}},
\bauthor{\bsnm{Clark}, \binits{T.T.}}
\bjtitle{Theor. Appl. Mech. Lett.}
\bvolume{7},
\bfpage{301}
(\byear{2017})
\end{barticle}
\endbibitem

\bibitem[\protect\citeauthoryear{B{\"o}hm-Vitense}{1997}]{bohm}
\begin{bbook}
\bauthor{\bsnm{B{\"o}hm-Vitense}, \binits{E.}}:
\bbtitle{Introduction to Stellar Astrophysics}.
\bsertitle{Stellar atmospgeres},
vol. \bseriesno{2}.
\bpublisher{CUP},
\blocation{Cambridge}
(\byear{1997})
\end{bbook}
\endbibitem

\bibitem[\protect\citeauthoryear{Levich}{1983}]{levich2}
\begin{barticle}
\bauthor{\bsnm{Levich}, \binits{E.}}
\bjtitle{Ann. N. Y. Acad. Sci.}
\bvolume{404}(\bissue{1}),
\bfpage{73}
(\byear{1983})
\end{barticle}
\endbibitem

\bibitem[\protect\citeauthoryear{Levich}{2009}]{levich1}
\begin{barticle}
\bauthor{\bsnm{Levich}, \binits{E.}}
\bjtitle{Concepts Phys.}
\bvolume{6}(\bissue{3}),
\bfpage{239}
(\byear{2009})
\end{barticle}
\endbibitem

\bibitem[\protect\citeauthoryear{Lee}{2000}]{lee1}
\begin{barticle}
\bauthor{\bsnm{Lee}, \binits{C.B.}}
\bjtitle{Phys. Rev. E}
\bvolume{62},
\bfpage{3659}
(\byear{2000})
\end{barticle}
\endbibitem

\bibitem[\protect\citeauthoryear{}{2008}]{lee2}
\begin{botherref}
C. b. lee and j. z. wu.
Appl. Mech. Rev.
\textbf{61},
030802
(2008)
\end{botherref}
\endbibitem

\bibitem[\protect\citeauthoryear{Lee and Jiang}{2019}]{lee3}
\begin{barticle}
\bauthor{\bsnm{Lee}, \binits{C.B.}},
\bauthor{\bsnm{Jiang}, \binits{X.Y.}}
\bjtitle{Phys. Fluids}
\bvolume{31},
\bfpage{111301}
(\byear{2019})
\end{barticle}
\endbibitem

\bibitem[\protect\citeauthoryear{Jiang et~al.}{2020}]{jiang}
\begin{barticle}
\bauthor{\bsnm{Jiang}, \binits{X.Y.}},
\bauthor{\bsnm{Lee}, \binits{C.B.}},
\bauthor{\bsnm{Chen}, \binits{X.}},
\bauthor{\bsnm{Smith}, \binits{C.R.}},
\bauthor{\bsnm{Linden}, \binits{P.F.}}
\bjtitle{J. Fluid Mech.}
\bvolume{890},
\bfpage{11}
(\byear{2020})
\end{barticle}
\endbibitem

\bibitem[\protect\citeauthoryear{Hopfinger and Browand}{1982}]{hopfinger}
\begin{barticle}
\bauthor{\bsnm{Hopfinger}, \binits{E.J.}},
\bauthor{\bsnm{Browand}, \binits{F.L.}}
\bjtitle{Nature}
\bvolume{295},
\bfpage{393}
(\byear{1982})
\end{barticle}
\endbibitem

\bibitem[\protect\citeauthoryear{Mozer et~al.}{2021}]{mozer}
\begin{barticle}
\bauthor{\bsnm{Mozer}, \binits{F.S.}},
\bauthor{\bsnm{Bonnell}, \binits{J.W.}},
\bauthor{\bsnm{Hanson}, \binits{E.L.M.}},
\bauthor{\bsnm{Gasque}, \binits{L.C.}},
\bauthor{\bsnm{Vasko}, \binits{I.Y.}}
\bjtitle{Astrophys. J.}
\bvolume{911},
\bfpage{89}
(\byear{2021})
\end{barticle}
\endbibitem

\bibitem[\protect\citeauthoryear{Temerin et~al.}{1982}]{temerin}
\begin{barticle}
\bauthor{\bsnm{Temerin}, \binits{M.}},
\bauthor{\bsnm{K}, \binits{C.}},
\bauthor{\bsnm{Lotko}, \binits{W.}},
\bauthor{\bsnm{Mozer}, \binits{F.S.}}
\bjtitle{Phys. Rev. Lett.}
\bvolume{48},
\bfpage{1175}
(\byear{1982})
\end{barticle}
\endbibitem

\bibitem[\protect\citeauthoryear{et~al.}{2004}]{zweben1}
\begin{barticle}
\bauthor{\bsnm{al.}, \binits{S.J.Z.}}
\bjtitle{Nucl. Fusion}
\bvolume{44},
\bfpage{134}
(\byear{2004})
\end{barticle}
\endbibitem

\bibitem[\protect\citeauthoryear{Zweben et~al.}{1982}]{zweben}
\begin{barticle}
\bauthor{\bsnm{Zweben}, \binits{S.J.}},
\bauthor{\bsnm{Liewer}, \binits{P.C.}},
\bauthor{\bsnm{Gould}, \binits{R.W.}}
\bjtitle{Bull. Am. Phys. Soc.}
\bvolume{27},
\bfpage{973}
(\byear{1982})
\end{barticle}
\endbibitem

\bibitem[\protect\citeauthoryear{Surko and Slusher}{1982}]{surko}
\begin{barticle}
\bauthor{\bsnm{Surko}, \binits{C.M.}},
\bauthor{\bsnm{Slusher}, \binits{R.E.}}
\bjtitle{Bull. Am. Phys. Soc.}
\bvolume{27},
\bfpage{937}
(\byear{1982})
\end{barticle}
\endbibitem

\bibitem[\protect\citeauthoryear{Meiss}{1984}]{meiss-book}
\begin{bbook}
\bauthor{\bsnm{Meiss}, \binits{J.D.}}:
\bbtitle{Statistical Physics and Chaos in Fusion Plasmas}.
\bsertitle{Nonequilibrium Problems in the Physical Sciences and Biology},
vol. \bseriesno{3}.
\bpublisher{Wiley--Blackwell},
\blocation{New Jersey}
(\byear{1984})
\end{bbook}
\endbibitem

\bibitem[\protect\citeauthoryear{Wang et~al.}{2020}]{debye-scale}
\begin{botherref}
\oauthor{\bsnm{Wang}, \binits{R.}},
\oauthor{\bsnm{Vasko}, \binits{I.Y.}},
\oauthor{\bsnm{Mozer}, \binits{F.S.}},
\oauthor{\bsnm{Bale}, \binits{S.D.}},
\oauthor{\bsnm{Artemyev}, \binits{A.V.}},
\oauthor{\bsnm{Bonnell}, \binits{J.W.}},
\oauthor{\bsnm{Ergun}, \binits{R.}},
\oauthor{\bsnm{Giles}, \binits{B.}},
\oauthor{\bsnm{Lindqvist}, \binits{P.-A.}},
\oauthor{\bsnm{Russell}, \binits{C.T.}},
\oauthor{\bsnm{Strangeway}, \binits{R.}}
Astrophys. J. Lett.
\textbf{889:L9}
(2020)
\end{botherref}
\endbibitem

\bibitem[\protect\citeauthoryear{Burch et~al.}{2016}]{mms}
\begin{barticle}
\bauthor{\bsnm{Burch}, \binits{J.L.}},
\bauthor{\bsnm{Moore}, \binits{T.E.}},
\bauthor{\bsnm{Torbert}, \binits{R.B.}},
\bauthor{\bsnm{Giles}, \binits{B.L.}}
\bjtitle{Space Sci. Rev.}
\bvolume{199},
\bfpage{5}
(\byear{2016})
\end{barticle}
\endbibitem

\bibitem[\protect\citeauthoryear{Das and Bora}{2024}]{mrid}
\begin{botherref}
\oauthor{\bsnm{Das}, \binits{M.}},
\oauthor{\bsnm{Bora}, \binits{M.P.}}
arXiv:2402.18478 [physics.plasm-ph]
(2024)
\end{botherref}
\endbibitem

\bibitem[\protect\citeauthoryear{Korteweg and de~Vries}{1895}]{kdv}
\begin{barticle}
\bauthor{\bsnm{Korteweg}, \binits{D.J.}},
\bauthor{\bsnm{Vries}, \binits{G.}}
\bjtitle{Phil. Mag.}
\bvolume{39},
\bfpage{422}
(\byear{1895})
\end{barticle}
\endbibitem

\bibitem[\protect\citeauthoryear{Peregrine}{1966}]{perigrene}
\begin{barticle}
\bauthor{\bsnm{Peregrine}, \binits{D.H.}}
\bjtitle{J. Fluid Mech.}
\bvolume{25}(\bissue{2}),
\bfpage{321}
(\byear{1966})
\end{barticle}
\endbibitem

\bibitem[\protect\citeauthoryear{Druyvesteyn and
  Penning}{1940}]{Druyvesteyn-1940}
\begin{botherref}
\oauthor{\bsnm{Druyvesteyn}, \binits{M.J.}},
\oauthor{\bsnm{Penning}, \binits{F.M.}}
Rev. Mod. Phys.
\textbf{12}(87)
(1940)
\end{botherref}
\endbibitem

\bibitem[\protect\citeauthoryear{Khusroo}{2019}]{murchana}
\begin{botherref}
\oauthor{\bsnm{Khusroo}, \binits{M.}}:
A study on plasma processes in astrophysical environment with an emphasis on
  magnetospheres and accretion disks.
PhD thesis,
Gauhati University,
http://hdl.handle.net/10603/291394
(2019)
\end{botherref}
\endbibitem

\bibitem[\protect\citeauthoryear{Frost and Phelps}{1962}]{frost}
\begin{barticle}
\bauthor{\bsnm{Frost}, \binits{L.S.}},
\bauthor{\bsnm{Phelps}, \binits{A.V.}}
\bjtitle{Phys. Rev.}
\bvolume{127},
\bfpage{1621}
(\byear{1962})
\end{barticle}
\endbibitem

\bibitem[\protect\citeauthoryear{Yoon et~al.}{2008}]{yoon}
\begin{barticle}
\bauthor{\bsnm{Yoon}, \binits{J.-S.}},
\bauthor{\bsnm{Song}, \binits{M.-Y.}},
\bauthor{\bsnm{Han}, \binits{J.-M.}},
\bauthor{\bsnm{Hwang}, \binits{S.H.}},
\bauthor{\bsnm{Chang}, \binits{W.-S.}},
\bauthor{\bsnm{Lee}, \binits{B.}},
\bauthor{\bsnm{Itikawa}, \binits{Y.}}
\bjtitle{J. Phys. Chem. Ref.}
\bvolume{37},
\bfpage{913}
(\byear{2008})
\end{barticle}
\endbibitem

\bibitem[\protect\citeauthoryear{Gubbins and Herrero-Bervera}{2007}]{elsasser}
\begin{bbook}
\beditor{\bsnm{Gubbins}, \binits{D.}},
\beditor{\bsnm{Herrero-Bervera}, \binits{E.}} (eds.):
\bbtitle{Encyclopedia of Geomagnetism and Paleomagnetism},
\bedition{1}st edn.
\bpublisher{Springer},
\blocation{Dordrecht}
(\byear{2007})
\end{bbook}
\endbibitem

\bibitem[\protect\citeauthoryear{Sano and Stone}{2002}]{sano}
\begin{barticle}
\bauthor{\bsnm{Sano}, \binits{T.}},
\bauthor{\bsnm{Stone}, \binits{J.M.}}
\bjtitle{Astrophys. J.}
\bvolume{577},
\bfpage{534}
(\byear{2002})
\end{barticle}
\endbibitem

\bibitem[\protect\citeauthoryear{Chen}{1974}]{chen-1974}
\begin{bbook}
\bauthor{\bsnm{Chen}, \binits{F.F.}}:
\bbtitle{Introduction to Plasma Physics and Controlled Fusion}
vol. \bseriesno{1},
\bedition{2}nd edn.
\bpublisher{Plenum Press, New York and London},
\blocation{New York}
(\byear{1974})
\end{bbook}
\endbibitem

\bibitem[\protect\citeauthoryear{Sagdeev}{1979}]{sagdeev}
\begin{barticle}
\bauthor{\bsnm{Sagdeev}, \binits{R.Z.}}
\bjtitle{Rev. Mod. Phys.}
\bvolume{51},
\bfpage{1}
(\byear{1979})
\end{barticle}
\endbibitem

\bibitem[\protect\citeauthoryear{Mott-Smth and Langmuir}{1926}]{mott}
\begin{barticle}
\bauthor{\bsnm{Mott-Smth}, \binits{H.}},
\bauthor{\bsnm{Langmuir}, \binits{I.}}
\bjtitle{Phys. Rev.}
\bvolume{28},
\bfpage{0727}
(\byear{1926})
\end{barticle}
\endbibitem

\bibitem[\protect\citeauthoryear{Allen}{1992}]{allen}
\begin{barticle}
\bauthor{\bsnm{Allen}, \binits{J.E.}}
\bjtitle{Phys. Scripta}
\bvolume{45},
\bfpage{497}
(\byear{1992})
\end{barticle}
\endbibitem

\bibitem[\protect\citeauthoryear{Shukla and Mamun}{2002}]{shukla}
\begin{bbook}
\bauthor{\bsnm{Shukla}, \binits{P.K.}},
\bauthor{\bsnm{Mamun}, \binits{A.A.}}:
\bbtitle{Introduction to Dusty Plasma Physics}.
\bpublisher{IOP},
\blocation{Bristol and Philadelphia}
(\byear{2002})
\end{bbook}
\endbibitem

\end{thebibliography}

\end{document}